# Optimization of parameters of a couple generator-receiver for a gravitational Hertz experiment

V.N.Rudenko

(Sternberg Astronomical Institute, MSU, Russia)

Abstract

Brief review of principal ideas, estimates and schemes proposed by Russian research groups in respect of the gravitational radiation generated and detected in the laboratory condition is presented. Analysis leads to conclusion that the more promising variant of the laboratory GW-Hertz experiment might be associated with power electromagnetic and acoustical impulsive or shock waves travelling and interacting in nonlinear optic-acoustical medium

One of the first consideration of the gravitational Hertz experiment in laboratory was done by J.Weber in his pioneer paper [1]. He found that a generated power might reach $10^{-13}\, erg/sec$ but remained to be much less the receiver sensitivity. Afterwards several different mechanical and electromagnetic schemes were analyzed [2-5].
Main conclusion was that in laboratory conditions of "slow motion" and "weak field" it would be extremely difficult to construct any effective gravitational generator and Hertz experiment seemed more as Gedanken figure then practical one. On the other hand an understanding that a gravitational radiation noise background at the typical AM-FM frequency range also must be negligibly small allowed to continue a speculation on possibility of a "gravitational transmission of information" even with weak power transmitters.
In this lecture we have the goal to estimate from a general point of view an upper limit of the power level reasonable for the terrestrial Hertz experiment taking into account principal physical restrictions, present and future technical feasibility. Our analysis is based mostly on the papers of Russian research groups published during the last quarter of past century, but similarly works were carried out in parallel by researchers in US and Europe approximately at the same time. All estimates below are made supposing a validity of General Relativity and any unconventional physics is not considered. We start with a conceivable GW-generator composed by a coherent group of mechanical oscillators and show that the requirement of acoustical resonance leads to the tendency of frequency decreasing which in its turn comes to enormously large scale of set ups even for a weak radiated power.
To improve the situation we subsequently consider following possibilities:
1) a refusing the resonance condition for a transition to high frequencies;
2) the increasing a number of coherent elementary radiators at the molecular scale;
3) an utilization some artificial medium with abnormally high sound velocity.
After all as a final conclusion we formulate our view to the more promising way of developing HFGW technique in the laboratory Hertz experiments at present.

## 1. GW-radiation of mechanical vibrators

Let's consider an elementary mechanical radiator consisting of elastically bound masses $m$ separated by a distance $l_0$ and having a vibration amplitude $\Delta l \ll l_0$. According to the "quadrupole formula" a total GW radiation power will be

$$P_g \cong (2G/45c^5)m^2 l_0^2 (\Delta l)^2 \omega^6 \qquad (1)$$

In practice such elastic dumb-bell is realized as a fundamental longitudinal mode of a solid rod with parameters $L, M$ so that $l_0 = (2/3)L$ and $m = (1/3)M$. Contradiction between requirements to have a "large vibration amplitude" and "high vibration frequency" leads to the condition of acoustic resonance excitation $l_0 = \lambda_a / 2$; this results in the following transformation of the expression (1)

$$P_g \cong (G/c^5)\rho_0^2 S_0^2 v_s^6 x^2 \qquad (2)$$

where $\rho_0, S_0$ are the density and cross-section of the rod, $x = \Delta l / l_0 = \Delta r_0 / r$ is the deformation amplitude, $v_s$ - velocity of sound; the formula (2) is valid also for higher overtones with odd numbers.

The conventional way of increasing the output power is associated with the coherent addition of fields of many radiators $n \approx \lambda_g / \lambda_a$, occupying a region with the scale $\sim \lambda_g / 2$; a cross section of such individual radiator $\sim (c/2\nu_g)^2$; (one could presents this square also as a linear composition of elementary cells with cross section $(\lambda_a / 2)^2$ so that a number of cells would be $n_s = (1/4)(\lambda_g / \lambda_a)^2$; then a total number of coherent radiators will be $N_r = n * n_s \approx (\lambda_g / \lambda_a)^3 \approx 10^{15}$). This multiple generator composed by $n$ elastic layers with the width $\lambda_a / 2$ and square $(\lambda_g / 2)^2$ has to produce increased power $\sim n^2$ so that the total radiation might be presented as

$$P \approx 10^{-3}(G/c^3)(\rho_0 v_s^2 x)^2 \lambda_g^4 \qquad (3)$$

In principle it demonstrates a benefit of frequency decreasing (so as it increases the coherent volume) but the payment would be undesirable expansion of the set up size $d \sim \lambda_g / 2$. It seems a maximum reasonable size of generator would be $d \sim 10\,m$, consequently $\nu_g \sim 10^7 \cdot Hz$; then for $\rho_0 \approx 5 \cdot g/cm^3$, $v_s \approx 5 \cdot 10^5 cm/\sec$ and for a largest allowed deformation $x \approx 10^{-3}$ one comes to the radiated gravitational power $P \leq 10^{-11} erg/\sec$. In principle this power could be increased at least one order of value using a line of identical set ups (from 3 up to10) with correspondent phase shifts necessary for the coherent power accumulation. The factor $10^{-3}$ in (3) reflects inhomogeneous density distribution over the generator volume. It tends to unit for the electromagnetic gravity generator [3] with homogeneous mass density $\rho_0 = \varepsilon/c^2$; in this case the electromagnetic energy density $\varepsilon$ is equivalent of the mechanical tension ($\rho_0 v_s^2 x$) in (3); thus the question is only which variable may be larger.

In the numerical example above the mechanical tension consists $4 \cdot 10^8 din/cm^2$ (close to the damage threshold $10^9 din/cm^2$) and corresponds to magnetic field amplitude $\sim 10^5 Gs$; however a maintaining such enormous field in the volume $\sim 10^3 m^3$ looks exotic technically; the more realistic value $\sim 10^4 Gs$ leads to the same numerical estimate of radiated power as it was for the mechanical model. Thus we conclude that the upper limit of laboratory generated gravitational radiation power in the radio range frequencies could not exceed $(10^{-10} - 10^{-11})\, erg/sec$; an adequate set up has to have almost an "industrial scale" and very large driving power $\sim 10^{10} W$ or even more.

Receiver for this GW-radiation might be thought as the usual resonance gravitational "bar detector" [2] but also composed by $N_d = 2500$ elementary sapphire cells (bars) with length $l_i = \mathbf{l}_a/2 \approx 0.1\ cm$ and diameter $\sim 30\ cm$, so that the total mass would consists $M \sim 10^5 g$. At the very low temperature $T = 10$ mK mechanical quality factor in principle might be so large as $Q_m = 10^{15}$. Then the minimum value of registering gravitational wave flux $I_\Sigma$ according to the formula

$$I_\Sigma = N_d^{-1} I_{\min} = \frac{2c^3}{\mathbf{p}^3 G}\, \frac{kT\mathbf{w}_0}{M v_s^{\,2} Q_m \mathbf{t}} \quad , \qquad (4)$$

(where $I_{\min}$ is the sensitivity of individual cell (bar), $\mathbf{w}_o = \mathbf{p} v_s / l_i$) results in $I_\Sigma \geq 10^{-13}\, erg/cm^2\, \sec$ for three months integration time $\mathbf{t}$.

Formula (4) describes a non coherent detection when signals of individual elementary bars undergo a simple addition. However in the laboratory Hertz experiment it is possible to apply a so called coherent detection using an adopted phase synchronization of the cells. Then $I_\Sigma = N_d^{-2} I_{\min}$ and the detectable flux would be $I_\Sigma \geq 10^{-16}\, erg/cm^2\, \sec$. This looks almost satisfactory for detection of the GW signal at the edge of the wave zone but a beaming of the radiated power remains to be desirable.

Thus at least on the level of numerical estimates the GW Hertz experiment seems as a feasible one but at the very low level of radiated and received power and with unreasonably large energetic supply expenses.

**2. Radiator with the "stretched acoustic wave"**

Investigation for developing these principal ideas on engineering level were carried out in Institute of Radio and Electronics (Rus. Ac. Sc., Moscow)) by the group of prof. M.Golubtsov [4]. Ferromagnetic electromechanical transducer was studied as GW- radiator and receiver. Total power radiated by the ferromagnetic rod ($l$- length, $d$ - diameter) in the magneto-acoustical resonance with harmonic number $n$ .is written as

$$P_g \approx \frac{G}{15\mathbf{p}^2 c^5}\, \mathbf{r}^2 \mathbf{w}^6 \left(\frac{dl}{n}\right)^4 \mathbf{x}^2 \qquad (5)$$

this formula is valid for $\mathbf{l}_g \geq l$ ; it shows that only $(1/n)$ part of the rod radiates effectively.(the radiation of other parts is self compensated).

Signal-noise ratio for the complete couple "generator – receiver" (supposing that the thermodynamic fluctuation of the receiver is the mains source of noise) at the border of the wave zone $(\mathbf{w}/c)r \approx 1$ might be presented in the form

$$\frac{S}{N} = P_g \left(\frac{P_r}{P_g}\right)\left(\frac{P_m}{P_r}\right)\frac{1}{2P_N} \approx \frac{15\mathbf{p}^{12}G\mathbf{r}^3 c^5}{16\mathbf{w}^6 kT\Delta f}\mathbf{x}^2 \left(\frac{d}{l}\right)^6 \left(\frac{\mathbf{l}_a}{\mathbf{l}_g}\right)^{15} N^2{}_r N^2{}_d \qquad (6)$$

where $P_r, P_m$ - are the gravitational and mechanical power in the receiver; $P_N$ - is the noise power; $N_r, N_d$ - numbers of coherent radiators and receivers.

The formula (6) demonstrates a dramatic role of "wave mismatch" for the "Hertz couple" in acoustical resonance: the factor $(\mathbf{l}_a / \mathbf{l}_g) \sim 10^{-5}$; besides it recommends to improve a geometry of radiator $(d/l) \sim 1$. A solution of the wave mismatch problem requires some "effective stretch" of the acoustical wave in the solid body. Under the fixed velocity of sound it means a refuse from the acoustic resonance condition and operation at frequencies much high the resonance frequency. At the same time it would be desirable to keep the deformation amplitude $\mathbf{x}$ not too small. As it follows from (1) factually one has to balance a contradiction between the frequency increasing and loose of oscillation amplitude by some optimal way.

In practice a realization of "bulk vibrations" in a solid body at $\mathbf{w} \gg \mathbf{w}_0$ is also non trivial problem. IRE group proposed to do it for transversal acoustical modes through conversion low frequency resonant elastic tensions to high frequency in the nonlinear magnetostriction media

Technically such generator might be realized in the following manner.
A coaxial cylinder from a magnetostriction material (coupling constant $\Lambda$) is placed inside the EM resonator. Alternative EM field of the resonator excites powerful acoustical oscillations of the fundamental transverse (tidal) mode $\mathbf{w}_2$ of the bar homogeneously along the bar axis. Another high frequency EM pump $\mathbf{w}_p \gg \mathbf{w}_2$ inserted into the resonator produces EM and elastic oscillations with combined frequencies $\mathbf{w}_{1,2} = \mathbf{w}_p \pm \mathbf{w}_2 \sim \mathbf{w}_p$. The key point in this interaction is the sidebands are born with the same space structure of the initial acoustical standing wave $\mathbf{w}_2$, i.e. the magnetostriction coupling provides a mechanism for forced bulk oscillations of the bar at very high frequencies; the amplitude of the forced oscillations can not be so large as the resonance one but this deficit is compensated by the frequency increasing. The conservation rule gives the following relation between wave numbers of gravitational, EM-pump and fundamental acoustical oscillations $n_{1,2} = |n_p - n_2| = n_g$. Numerical estimation with ferromagnetic parameters:
$Y \sim 2\cdot 10^{11} N/M^2$, $\mathbf{m} = (\Delta\Lambda/\Lambda) \sim 0.8$, $H \sim 10E$, $\mathbf{s} \approx 1.6\cdot 10^{-5}Y$, leads to deformation $\mathbf{x} = (\Delta r/r) \sim 2\cdot 10^{-5}$ for acoustical mode $\mathbf{w}_2$ and only $\mathbf{x} \approx 2\cdot 10^{-14}$ for the sideband frequency $\mathbf{w}_1 \sim 10^8 s^{-1}$ if the scale of the rode has the order $l \approx d \sim 1M$.
Nevertheless a substitution in the (6) $\mathbf{w} = 2\mathbf{p}10^8 s^{-1}$, $(\mathbf{l}_a / \mathbf{l}_g) = 1$, $(\mathbf{s}/Y) \sim 10^{-3}$, $\mathbf{r} \approx 8.3 g/cm^3$, $T = 4.2K$, $\Delta f = 0.01 Hz$, - forecasts the $(S/N) \sim 10^{-7}$ for single couple "generator – receiver". It is clear that using $N_r = N_d = 100$ couples one can reach the threshold $(S/N) \sim 1$. As a whole the idea of equivalent "acoustical wave starching" to increase efficiency of the GW-generation in spite of its reasonable character was never used in practice yet.

**3. Nonlinear interaction of travelling waves.**

In the scheme of Hertz experiment with standing waves described above the benefit of frequency increasing was paid by the loss of mechanical vibration amplitude.
It is clear from the key formulae (1), (2), (6) that for to keep the resonance condition (i.e. maximum vibration amplitude) one has to decrease the mass (scale) of elementary radiator taking into account a possibility of using a natural mass-quadruple at the molecular level. The deficit of mass in this case might be covered by increasing the total number of coherent micro radiators and detectors $N_r, N_d$. It is obvious the maximum density of cells can not exceed the value A= $10^{22} - 10^{24}$ per $cm^3$. After this argumentation it is reasonable to seek a solution of the problem effective GW generation on the way of collective excitation of molecular oscillations at optical frequencies taking place in a nonlinear optic-acoustical media. One has to choose a molecular with largest quadruple momentum; an instrument for their excitation might be a powerful optical laser beam; a coupling of optical, acoustical and gravitational waves must be provided by nonlinear optical permeability of the media. Accumulation of the GW power as well as its beaming might be provided by the synchronism conditions typical for interaction of waves travelling through nonlinear optical media.

These ideas had been studied in the Joint Institute of Nuclear Research (JINR, Dubna) at the end of the past century in the group of prof A.Pisarev and P.Bogolyubov [5,6]. In particular they considered as a proper media the matrix crystal artificially prepared from the molecular hydrogen $H_2$ dissolved into argon $Ar$ (1:10) and cooled up to helium temperature for a phase transition into a solid state [5].

Phenomenology of the process of coherent transformation parametrically coupled in a nonlinear media optical and gravitational waves might be briefly described as it follows. Let an electromagnetic polarization of some nonlinear optically transparent media is $P = \boldsymbol{a}E$, where the electrical permeability is a function of an electrical field amplitude $E$ and metric perturbation (or deformation) $h$, i.e. $\boldsymbol{a} = \boldsymbol{a}(E, h)$. Under the action of travelling electromagnetic pump waves an induced variable part of polarization might be presented as $\Delta P = (\partial \boldsymbol{a}/\partial q)|_{q_0} < q > (E_1 + E_2)(1 + \boldsymbol{b}h) + ...$ where $q$ is molecular center mass coordinate; parameters $(\partial \boldsymbol{a}/\partial q)|_{q_0}, \boldsymbol{b}$ define the variance of polarization under the action of electrical fields and mechanical quadruple deformations of molecular. Due to this two traveling electromagnetic pumps with wave parameters $\boldsymbol{w}_1, \vec{k}_1, \vec{E}_1$ and $\boldsymbol{w}_2, \vec{k}_2, \vec{E}_2$ will produce the wave of molecular quadruple deformations which in its turn will generate the gravitational wave $\Omega_g, \vec{k}_g, \vec{h}$. The conservation law leads to the synchronism conditions $\Omega_g = 2(\boldsymbol{w}_1 - \boldsymbol{w}_2)$, $\vec{k}_g = 2(\vec{k}_1 + \vec{k}_2)$, which means that for effective interaction corresponding waves must travel along the definite directions with mutual angles depending on refractive indexes $n_1, n_2$ and ratio of frequencies $\boldsymbol{w}_1/\boldsymbol{w}_2$.
A final formula for the estimate of GW radiation power looks approximately as [5]

$$P_g \approx 2\boldsymbol{p}^5 (\partial \boldsymbol{a}/\partial q)|^4_{q_0} \frac{G N_r^2 P_1^2 P_2^2 L^2}{m^2 n_1 n_2 c^7 S \boldsymbol{g}_1 \boldsymbol{g}_2} \quad (7)$$

$m$ - is molecular mass $H_2$, $\boldsymbol{g}_1 \sim \boldsymbol{g}_2 \approx 10^7 \sec^{-1}$ - widths of optical resonances, $S, L$ -are the beam cross section and coherent length. A substitution the following numerical parameters [5]: $P_1, \sim P_2 \sim 3 \cdot 10^9 W$ (a laser pulse with duration $\sim 10^{-6}$ sec), $(\partial \boldsymbol{a}/\partial q|_q \sim 3 \cdot 10^{-15} cm^2$ (taken from combined light scattering experiments), $N_r \approx 10^{23}$, $S = 1\ cm^2$, $L = 1\ M$ results in the estimate $P_g \sim 0.1\ erg/\sec$.

GW detector might be arrange according to the reversed scheme: the gravitational wave and power EM-pump will produce the electromagnetic "signal wave" at the combined frequency $\boldsymbol{w}_s$ as it was marked above. Calculation of the Dubna group [5,6] forecast the detection efficiency approximately $P_s \sim 100 h \boldsymbol{n}$ per $P_g = 1\ erg/\sec$ (one hundred signal photons per one erg/sec gravitational power) under the EM-pump power 1 $GWt$, resonance width $\boldsymbol{g} \sim 10^7 \sec^{-1}$ and frequency shift $(\boldsymbol{w}_p - \boldsymbol{w}_s) \approx 3\boldsymbol{g}$ i.e. enough to be registered by modern photo detectors.

At practice this variant of the Hertz experiment also never been realized, although it looks very promising as a pulse type GW communication at the optical frequency range.

## 4. Pulse GW generation by micro explosions.

The formula (2) for GW-generator in acoustical resonance exhibits the following law of radiation power $P_g \sim \boldsymbol{r}^2 v_s^6$. It stimulate a search for medium with increased values of the density and especially the sound velocity. Solid bodies have the maximum magnitudes on order of $\boldsymbol{r} \sim 10\ g/cm^3$, $v_s \sim 10^6\ cm/\sec$. So the only possibility to get more is to deal with dynamically compressed matter. For example if there is some gas with the polytropic state equation $p = C\boldsymbol{r}^{\boldsymbol{g}}$, then one gets for the sound velocity $v_s = (p/\boldsymbol{r})^{1/2} \sim \boldsymbol{r}^{(\boldsymbol{g}-1)/2} \sim \boldsymbol{r}^{1/3}$, it means in such media the GW radiation power will rise after the compression according to law $P_g \sim \boldsymbol{r}^4$ !. It leads to the idea of using an "implosion" or "inside directed burst process", i.e. a specially organized directed explosion with the matter moving to its internal region increasing the density. These ideas were developed by prof. V.Belokogne [7] who calculated the GW-radiation from multi layer targets compressed by a very powerful and short laser pulse.

In particular he considered a model of so called "clapping book" in which a packet of parallel plates from a very thin aluminum foil irradiated by laser beams from opposite sides. Enormous light pressure compresses plates converting its matter in a high energy plasma with shock waves travelling and bouncing "to and back" many times inside the target. Just this shock wave dynamics produces a short pulse GW radiation with power $P_g \sim (G/c^5)(\boldsymbol{h} d \mathrm{E}_{mec}/dt)^2$, where $\mathrm{E}_{mec}$ is the energy of mechanical movements, $\boldsymbol{h}$-its quadruple fraction. Numerical parameters of the model used in [7] were the plate mass- $m_i \sim 10^{-3}\ g$, radius- $r \sim 10^{-2}\ cm$, width- $d \sim 10^{-3}\ cm$ number of layers- $n \sim 10^2 - 10^3$; laser pulse power- 10 $MJ$, duration- $\Delta t \leq 10^{-9}\ \sec$.

Calculation in [7] gives the effective increasing of sound velocity $v_s \geq 10^8\ cm/\sec$ and density $\boldsymbol{r} \sim 10^3 - 10^5\ g/cm^3$. under the pressure $p \leq 10^{10} - 10^{11}\ atm$. Expected pulse of output GW radiation at frequencies $\boldsymbol{w}_g \sim 10^{13} - 10^{15}\ \sec^{-1}$ has to achieve $P_g \leq 1\ W$ inside $\Delta t \sim 10^{-12} - 10^{-14}\ \sec$ if the quadruple part of mechanical power would be $\boldsymbol{h} d\mathrm{E}_{mec}/dt \leq 10^{26}\ W$. Relatively large output power compare with the previous examples of GW generators is explained by relativistic character of the

source,-superdense high temperature plasma with relativistic velocities of particles and shock waves inside of it. So the "third derivative of the mass quadruple momentum" reaches large value despite of a small total mass of the target (in some sense we have here just the "jerk generator" mentioned by prof. R.Baker at this conference [8])

Although this idea looks very attractive and might be studied and technically realized as some by-product in the experimental research of a Laser Controlling Nuclear Fusion [9] it has a lack of absence a clear understanding of a possible detection mechanism for such extremely short very high frequency GW bursts. Nevertheless its study has to be continued for a developing a more regroups theory, technical design, proper laser source etc.

### 5.Conclusions.

After this very brief analysis of principal ideas and schemes more or less studied in Russia during of last thirty years the author should like formulate his personal opinion concerning a perspective way of solving the "GW Hertz laboratory experiment problem".
1. It was shown that Hertz-experiment on elastic standing waves using the "acoustical wave stretching technique" is feasible in principle at radio frequencies $(10-100)MHz$
However it would deal with a very weak generated GW power and requires a too large energy supply for continuos operation (in fact an middle scale electrical plant is needed). The scale of the set up has to exceed probably the standard laboratory facilities and also it would be difficult to find any reasonable application and development such technique besides a simple demonstration purpose and a prove the principal possibility. So it is unlikely this variant would be taken for future experiments.
2. GW Hertz experiment on travelling electromagnetic and acoustic (quadruple)waves interacting in nonlinear medium looks enough reasonable and promising at the range of optical frequencies $(10^{14}-10^{15})Hz$. It could provide a generation of short GW pulses $(10^{-6}-10^{-7})$ sec. with the high frequency carrier which might by used afterwards for a communication, new type of spectroscopy etc. Factually this method is the last development of early theoretical ideas of M.Gertzenstein [10], U.Kopvillem [11] and D.Galtsov [12] concerning a mutual conversion electromagnetic and gravitational wave in different media. Now it seems that modern achievements of the laser technique and nonlinear optics provide an adequate instrumental base to start at present such type of experiments. The only problem to be addressed seriously is a careful selection of proper nonlinear optic-acoustic material with large non linearity, enough quadruple molecular dynamics and small dissipation. Thus as we believe this type of Hertz experiment can be considered at present as a more profitable way for R&D research.
3. GW generation by the laser implosive matter compression certainly might produce extremely short $\Delta t \leq 10^{12}$ sec, powerful $\geq 10\,W$, GW pulses at optical and X range frequencies. However at present it is difficult to propose any detection principle for such pulses with duration shorter the typical relaxation times of atomic and molecular structures. For this reason now such version of the Hertz experiment needs to be studied deeply at first on a principal level then in technical details

**Acknowledgment**
First of all author should like express his sincere gratitude to his Russian colleagues prof. N.A.Armand, M.G.Golubtsov (IRE Rus.Ac.), A.F.Pisarev (JINR) V.A.Belokogne (MSU) for fruitful discussions and a very valuable assistance provided an access to archives materials of their institutions.

Also author passes his thanks to prof R.M.L.Baker, Jr for his extraordinary efforts in organizing HFGW conference and also for a stimulation encourage for a writing this paper.